\documentstyle[aps,epsfig]{revtex}
\begin{document}

\draft
\title{Influence of the left-handed part of the neutrino mass matrix on
the lepton number violating $e^-e^- \rightarrow W^-W^-$ process}

\author{P.~Duka, J.~Gluza, M.~Zra{\l}ek} 

\address{Department of Field Theory and Particle Physics 
Institute of Physics, University of Silesia 
Uniwersytecka 4, PL-40-007 Katowice, Poland
E-mails: duka,gluza,zralek@us.edu.pl} 
\vspace{1cm}
\date{\today}
\maketitle
\begin{abstract} 
Influence of the neutrino mass submatrix $M_L$ on the $e^-e^-
\rightarrow W^-W^-$ process is discussed. Taking into account various
possible CP signatures of heavy neutrinos it is shown that, in some cases,
nonzero $M_L$ substantially changes predictions for maximum possible values
of the $e^-e^- \rightarrow W^-W^-$ cross 
section. A direct role of the
$\omega^2$ parameter (coming from neutrinoless double 
beta decay) is clarified. The consequences of 
doubly charged Higgs particles $( \delta^{--}$) with resonances even far 
away from energies of the future linear lepton collider 
($\sqrt{s}=0.5 \div 1$ TeV) are studied. 
\end{abstract}
\vspace{0.5 cm}


\section{Introduction}

The $e^-e^-$ option of the future $\sqrt{s}=0.5 \div 2$ TeV linear collider 
is a very interesting area for investigating new physics \cite{snow}.
Processes like $e^-e^- \rightarrow l_i^-l_j^-,\;W^-W^-$ ($l_{i(j)}=e, \mu,\tau$)
could be studied indicating that lepton number (flavour or total) is not a global symmetry of the
electroweak interactions. In this paper we would like to examine the $e^-e^- 
\rightarrow W^-W^-$ process. This reaction violates the total lepton number by
two units, $\Delta L=2$. Its potential importance and hopes connected with 
it are based on two facts. First, the SM background is very small and
under control \cite{kol} so with the planed luminosity 
10 $fb^{-1}/year$ \cite{snow} the cross section as small as 0.1 fb
could give visible effect.
Second, its occurrence would indicate that there exist 
massive neutrinos of Majorana type. These neutrinos 
must be heavy (with masses $M_N>M_Z$) as known neutrinos cannot give any
substantial signal  \cite{gl6}. Many papers have 
been devoted to this process during the last decade \cite{rizzo,mink1,bel,eeww}.
For the first time this reaction was proposed and examined in 1982 by Rizzo
\cite{rizzo}. Additional interest has come with the paper \cite{mink1} where it
was
shown that the process is enhanced for heavy neutrino masses in the vicinity of 
the collider's c.m. energy. Then the optimism 
was revised in \cite{bel} where constraints on heavy neutrinos coming from
neutrinoless double-$\beta$ decay have been taken into account. It has been shown
that an observable signal requires fine-tuning among different heavy neutrino
couplings. However, as it has been shown in \cite{gl7}, these cancelations
can be in a natural manner connected with CP parities of heavy neutrinos.
All other papers cited in \cite{eeww} give many interesting
details connected with the process. 

This paper brings another such a detail which can however appear to be crucial 
for the magnitude of the cross section. In the last paper concerning the 
$e^-e^- \rightarrow W^-W^-$ where all relevant constraints on the heavy neutrinos
have been  taken into account \cite{gl7} 
we have assumed that the neutrino mass submatrix $M_L$ generated
by left-handed neutrinos

\begin{equation}
M=\left( \matrix{ M_L & M_D \cr
      M_D^T & M_R } \right),
\end{equation} 

is exactly zero ($M_D$ is the submatrix of Dirac type masses, $M_R$ is the submatrix of 
right-handed Majorana masses). This means that we have considered the class
of models beyond the standard where only right-handed neutrinos were
introduced. Then, to get an observable magnitude of the 
cross section, at least 3 heavy neutrinos  with appropriate CP 
signatures and masses were  necessary. However, there are models where $M_L$
does not vanish. Such nonzero 
$M_L$ changes the relations which restrict the space of parameters of possible
(i.e. allowed by experimental data) heavy neutrinos couplings and masses.

The full phenomenological discussion of non-zero 
$M_L$ has been given lately \cite{gl10} 
in the 
context of  heavy neutrino production in $e^-e^+$ ($e^-e^+
\rightarrow \nu N$) and $e^-\gamma$ ($e^-\gamma \rightarrow W^-N$)
reactions.
Here we will  restrict ourselves to two nonstandard models
with possible nonzero $M_L$.
The Standard Model with both additional right handed neutrinos (RHS) and Higgs
triplets and the Left-Right
symmetric model (LR). Details of these models can be found in literature
(e.g. in \cite{gl6,gl5}). 
As we are going to find the largest possible values of the cross section we
consider  models where CP is conserved in the lepton sector.

\section{The influence of $M_L$ on $e^-e^- \rightarrow W^-W^-$.}

The leading helicity amplitudes for the 
$e^-e^- \rightarrow W^-W^-$ process can be written in
the following, simplified, way 
\begin{equation}
M = \sum\limits_a \left\{ K_{ae}^2m_a \left[ f_t(m_a)+f_u(m_a)+
f_s^L \right]+\left( K_R \right)_{ae}^2 m_a f_s^R \right\}.
\end{equation}

The matrices $K,K_R$ are part of the unitary matrix $U=\left( K^{\ast},K_R
\right)^T$ which diagonalizes the $6 \times 6$ neutrino mass matrix M in Eq.(1),
the index L(R) is connected with the left (right) doubly charged Higgs
particle which is exchanged in the s channel. For details see e.g. \cite{gl6},
\cite{gl5}.

The sum in Eq.(2) runs over all light $(\nu)$ and heavy (N) neutrinos.
Let's note that the kinematical factors in t and u channels $f_{t,u}$ 
depend on $m_a$, but
the $f_s^{L(R)}$ ones in s channel do not.
First we will examine the t and u channels  assuming only that the
influence of the s channel is negligible (heavy $\delta_{L,R}^{--}$).
At the end we will comment on the effect of non-zero $M_L$ on the 
s-channel contribution. 

The experimental bounds on the elements of the matrix, $K_{\nu e}$ and
$K_{Ne}$, describing the mixing of electrons with light and heavy
neutrinos can be summarized as follows

\begin{equation}
\sum\limits_{N(heavy)} \left| K_{Ne}^2 \right| \leq \kappa^2 =0.0054,
\end{equation}

\begin{equation}
\sum\limits_{\nu (light)} \left| K_{\nu e}^2 m_{\nu} \right| \leq \kappa^2_{light}
=0.65\;eV,
\end{equation}

\begin{equation}
\left| \sum\limits_{N(heavy)} K_{Ne}^2\frac{1}{m_N} \right| \leq
\omega^2 =6 \cdot 10^{-3} \div 5 \cdot 10^{-5} TeV^{-1}.
\end{equation}
 
The first relation (Eq.(3)) comes from low energy experiments \cite{low}, the 
other ones can be derived from the fact that
neutrinoless double-$\beta$ decay $(\beta\beta)_{0\nu}$ 
has not been detected yet\footnote{As we can see there exist large 
discrepancies in the limit on
$\omega^2$. For arguments on lower (upper) limits, see 
\cite{gm} (\cite{bb}).}. 

Diagonalization of the matrix (1) together with (4)
yields to the following relation ($m_L={(M_L)}_{\nu_e \nu_e}$)

\begin{equation}
\left| m_L- \sum\limits_{N}K_{Ne}^2m_N \right| < \kappa_{light}^2.
\end{equation}

\hspace*{-1cm}
However, $\kappa^2_{light}$ is very small and can be neglected, then from
(6) we get 
\begin{equation}
\sum\limits_N m_NK_{Ne}^2 = m_L.
\end{equation}

Similar to the analysis given in \cite{gl7,gl10} let us discuss the 
influence of $m_L$
on the magnitude of the cross section for different CP parities of heavy
neutrinos. 

If we have only one heavy neutrino state (or more but with the
same CP parities) then from (3),(5) and (7) we get restrictions on $m_L$
\cite{gl10}

\begin{equation}
0 \leq m_L \leq min(\kappa^2 M, \omega^2 M^2),
\end{equation}

where M is the mass of the lightest of heavy neutrinos.
It gives for instance $m_L \leq 5 \cdot 10^{-4}$ GeV for M=100 GeV
($\omega^2=5 \cdot 10^{-5}\;TeV^{-1}$).
For the above values of $m_L$ the mixing angle $K_{Ne}$ is limited to
\cite{gl10}
\begin{equation}
K_{Ne}^2 \leq min (\omega^2 M, \kappa^2 ).
\end{equation}

Fig.1 shows the maximum value of the cross section $\sigma(e^-e^-
\rightarrow W^-W^-)$ where the parameters are restricted by
relation (9) for three different
values of $\omega^2$ and $\sqrt{s}=0.5(1)$ TeV. 
We can see that for various
$\omega^2$ there are different masses $M_0$ for which the cross section
reaches maximum value, e.g. $M_0 \simeq 1(100)$ TeV for 
$\omega^2=6 \cdot 10^{-3}
\;(5 \cdot 10^{-5}) \;TeV^{-1}$. For $M \leq M_0$ the maximum value of the
cross section increases with increasing M ($(K_{Ne})_{max}^2$ in Eq.(9) increases), 
for masses larger than $M_0$ the cross section decreases with M
($(K_{Ne})_{max}^2=\kappa^2=$const). We can see that only for 
$\omega^2 > 5 \cdot
10^{-4}\; TeV^{-1}$ and $\sqrt{s} \geq 1$ TeV there is a small region of
masses where $\sigma_{max} > 0.1$ fb. 
If there is only one heavy neutrino or more but with the same CP parities
then the value of $\omega^2$ crucially determines $\sigma_{max}$.
Much effort is devoted to find the bound on $\omega^2$ parameters \cite{gm}.

For the case of two heavy neutrinos with opposite CP parities we get the 
following inequalities ($K_{N_1e}=x_1,\;K_{N_2e}=ix_2,\;m_1=M,\;m_2=AM$)

\begin{eqnarray}
x_1^2+ \left|\frac{x_1^2}{A}-\frac{m_L}{AM} \right| & \leq & \kappa^2, \\
\left| x_1^2 \left( 1-\frac{1}{A^2} \right) +
\frac{m_L}{A^2M} \right| & \leq & \omega^2 M.
\end{eqnarray}

When $m_L=0$, to remove the bound given by $\omega^2$ (Eq.(11)), we have to
assume that two neutrinos are almost degenerate: $A \rightarrow 1$. But then
we have practically one Dirac neutrino (two Majorana neutrinos with opposite
CP values) and the cross section approaches zero.
This  was actually shown in \cite{gl7} where the $m_L=0$ case was examined. 

However, for $m_L \neq 0$ situation is different. The inequalities (10) and
(11) can be satisfied only for the confined region of $m_L$ \cite{gl10}

\begin{equation}
-max(AM\kappa^2,(A-1)M\kappa^2+A^2\omega^2M^2) \leq m_L \leq
min(\omega^2M^2, \kappa^2 M). 
\end{equation}

Positive values of $m_L$ are strongly restricted but the space of negative
$m_L$ values is wider and depends on the values of M and A. In Fig.2
we plot the results for $m_L=-1(-3,-5)$ GeV and A=5 as a function of neutrino
mass. 
As we can see lines start from different masses.
This is because Eq.(12) must holds. Similar results can be obtained for
larger spectrum of A $(=3 \div 15)$. For positive $m_L$ the situation is
similar to the case with $n_R=1$ (compare Eqs.(8),(12)). 
Results given in  Fig.2  describe also the case
of three  heavy neutrinos with the following CP signatures 
$\eta_{CP}(N_1)=-\eta_{CP}(N_2)=-\eta_{CP}(N_3)$. Then two heavy neutrinos
$(N_2,N_3)$ contribute in the same way to the amplitude (Eq.(2)) and can be
effectively treated as one.

The last quantitatively distinguishable possibility which is left 
for three heavy
neutrinos is  the case $\eta_{CP}(N_1)=\eta_{CP}(N_2)=-\eta_{CP}(N_3)$. 
Then  initial inequalities (3,5) and (7) are satisfied if $m_L$ is
confined to the following region ($m_1=M,\;m_2=AM,\; m_3=BM$) \cite{gl10}

\begin{eqnarray}
-min\left\{ BM\kappa^2, max \{ B^2\omega^2M^2, (B-1)M\kappa^2+B\omega^2M^2 \} 
\right\} \leq m_L && \nonumber \\
\leq min\{ AM \kappa^2,(A-B)M\kappa^2+AB\omega^2M^2 \}.
\end{eqnarray}

By fixing B=10 and M=100 GeV for different values of A we have found
mixing
angles $K_{N_1e}=x_1,\;K_{N_2e}=x_2,\;K_{N_3e}=ix_3$ such that the cross
section is maximal. The result is given in Fig.3 for $\sqrt{s}=1$ TeV. For
larger masses ($M>100$ GeV) $\sigma_{max}$ decreases, e.g. for M=200 GeV,
$\sigma_{max} \leq 4$ fb.
Let us note that the largest results are possible
for  large A and then we can always find
a space of allowed mixings for which $\sigma_{max} \simeq 9$ fb
independently of $m_L$.
Similar plots can be made for 
other energies $0.5\;TeV \leq \sqrt{s} \leq 2$ TeV with a result
$\sigma_{max} \leq 1(25,40)$ fb and $\sqrt{s}=0.5(1.5,2)$ TeV, respectively
(see \cite{gl7} for the $m_L=0$ case). 

Finally, in Fig.4  we describe the s-channel contribution to the $e^-e^-
\rightarrow W^-W^-$ process. We present the contribution of two doubly
charged Higgs particles $\delta_L^{--}$ and
$\delta_R^{--}$ which exist for example in the LR model\footnote{Other aspects of doubly charged Higgs physics at an
$e^-e^-$ collider can be found in  \cite{gun}}. 
Masses of the 
$\delta_{L,R}^{--}$ particles depend on $M_{W_2}$  \cite{gl3}
and for $M_{W_{2}}$=1
TeV we have (without fine tuning between parameters in the Higgs potential) 
$M_{\delta_{L}^{--}} \simeq 1600$ GeV and $M_{\delta_{R}^{--}} 
\simeq 3000$ GeV. As $m_{\delta_R^{--}} >> m_{\delta_L^{--}}$ the effect of
$\delta_R^{--}$ is negligible. In such circumstances our considerations are 
also valid for the SM enlarged by additional Higgs triplet and right handed
neutrinos.
Let us note that the contribution of the $\delta_L^{--}$ resonance to the
helicity amplitudes (Eq.(2))
 is directly proportional to $m_L$
($m_L=\sum\limits_a K_{ae}^2m_a$, see e.g. \cite{gl5}) and is invisible if
only light neutrinos exist. If we take $m_L=0$ then the
resonance disappears (solid line in Fig.4). If however $m_L \neq 0$ then 
its effect can be large. 
This is shown in Fig.4 where we take doubly charged Higgs' widths to be
$\Gamma_{\delta_{L,R}^{--}}=\Gamma_{M_W}M_{\delta_{L,R}^{--}}/M_W$. 
Lines on this Figure present the cross sections for the case when
all heavy neutrinos have the same CP eigenvalues.
 As it has already been discussed, 
$\sigma_{max}$ depends strongly on $\omega^2$ in this case. We take
$\omega^2=5 \cdot 10^{-5}\;TeV^{-1}$, so t and u channels contributions
to the cross section are very small (see Fig.1). 
This means that the large cross sections in Fig.4, even for energies far away 
from resonance region,
are due to the $\delta_L^{--}$ resonance.
For example for $m_{L}=5$ GeV and $\sqrt{s}=1$ TeV the 
$\sigma_{max}\simeq 40$ fb  and the effect is caused almost exclusively by 
$\delta_L^{--} (1600)$ Higgs resonance.
As the contribution of $\delta_R^{--}$ to the cross section  is not
proportional to $m_L$ its effect can be large even for $m_L=0$ especially if
its mass is around the C.M. energy. This case has been considered in [16].

This is shown explicitly in Fig.5 where
s-channel contribution to the $e^-e^- \rightarrow W^-W^-$ process
as function of energy is presented. 
To extract the effect of the
$\delta_L^{--}$ resonance we compare the cross section $\sigma_{max}$ for
t and u channels only (short-dashed line) with the total cross section
where t,u and s channels are added altogether (long-dashed lines) for 
$m_L=1$ GeV and $\eta_{CP}(N_1)=\eta_{CP}(N_2)$. 
We can see the huge influence of $\delta_L^{--}$ resonance on 
the total cross section. Even for very high mass of $\delta_L^{--}$
($M_{\delta_{L}^{--}}=2000$ GeV, $\sqrt{s}=1$ TeV) $\sigma_{max}$ 
is above the "discovery limit". 

Solid lines in Fig.5 describe  another case with
$\eta_{CP}(N_1)=-\eta_{CP}(N_2)$. The upper one corresponds to the full 
cross section (s,t,u channels),
the lower one is for a cross section without the s channel. As we can see
for $m_{L}=-1$ GeV,
contributions of the s and t+u channels are now comparable. 
The influence of $\delta_{L}^{--}$ on the cross section
depends on the $\delta_{L}^{--}$ mass and width, and the value of the
$m_{L}$ parameter. For the same mass and the same width of $\delta_{L}^{--}$ 
its  contribution to $\sigma_{max}
(e^-e^- \rightarrow W^-W^-)$ can be very small, comparable or much bigger
than the t+u channels' part depending on the value of $m_L$.

\section{Conclusions}

We have analyzed the predictions for maximum possible value of the
$e^-e^- \rightarrow W^-W^-$ cross section in models with nonzero $m_L$.
If we have only one heavy neutrino or more but with the same CP parities 
then the value of
$\omega^2$ is crucial for the maximum of the cross section and 
$m_L$ does not have any visible influence. For the smallest value of
$\omega^2$ ($\leq 5 \times 10^{-5}\;\mbox{\rm TeV}^{-1}$) predicted by some
existing estimations (e.g. \cite{bb}) the cross section $\sigma_{max}$ is too
small to be measured in the future $e^-e^-$ linear colliders unless 
$\delta_{L,R}^{--}$ exist in the model.

However, for all other cases, the
$m_L \neq 0$  changes substantially 
the $e^-e^- \rightarrow W^-W^-$ cross section. 
Negative $m_L$ values move the limits on experimentally allowed neutrino 
mixings and masses. If there are 
two heavy neutrinos with opposite CP parities 
(or any number of them but with the lightest one having opposite CP parity 
with respect to all other ones)
the value of $\sigma_{max}$ can be substantial,
much above the background level (e.g. for M=150 GeV, A=5, $\sigma_{max}
\simeq 7$ fb).

In another configuration of CP parities of heavy neutrinos
($\eta_{CP}(N_1)=\eta_{CP}(N_2)=-\eta_{CP}(N_3)$) 
the largest $\sigma(e^-e^- \rightarrow W^-W^-)$ is obtained for $m_L=0$. 

The most dramatic influence of the non-zero $m_L$ on $e^-e^- \rightarrow
W^-W^-$ is connected with the $\delta_L^{--}$ resonance. For 
$m_L=0$ the contribution of this
resonance to the process disappears. The $m_L \neq 0$ values cause  
the $\delta_L^{--}$ to give  large contribution 
even far away from on-peak energies. 
The contribution of the $\delta_R^{--}$ to the cross section does not
depend on the $m_L$ value.

\section*{Acknowledgments}
This work was supported by Polish Committee for Scientific Research
under Grant No. PB659/P03/95/08 and University of Silesia internal Grant.
\newline
J.G. also appreciate the fellowship of the Foundation for Polish Science.



\begin{figure}[h]
\epsfig{file=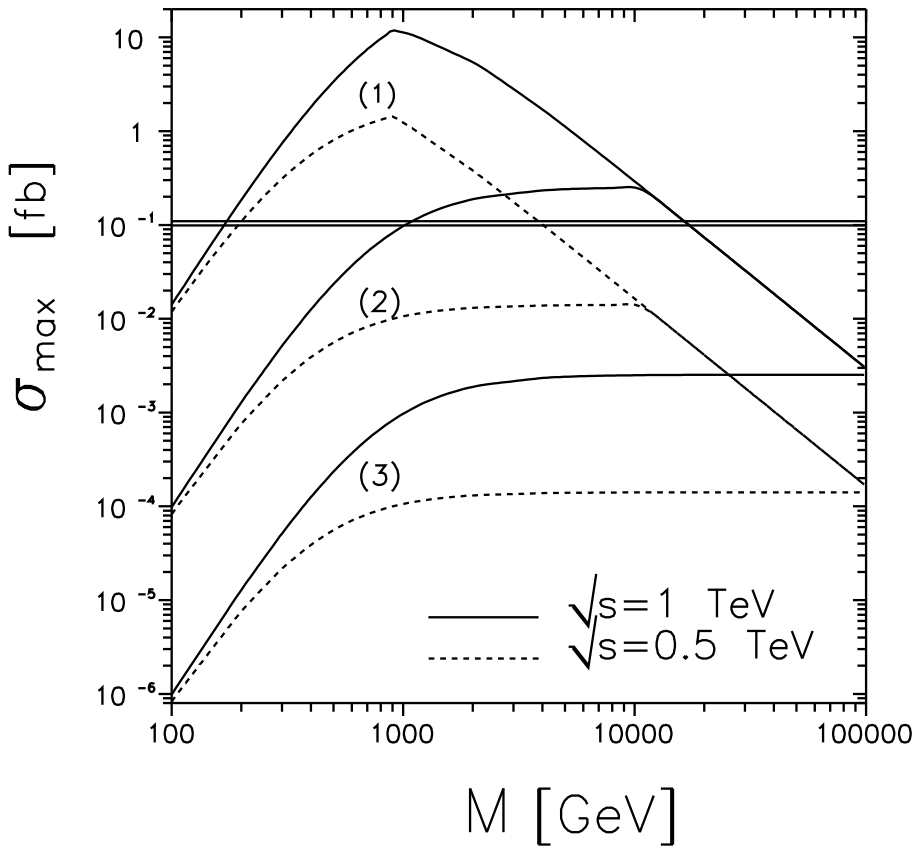,width=13cm,angle=90}
\caption{The largest cross section for the 
$e^-e^- \rightarrow W^-W^-$ process with
any number of heavy neutrinos with the same CP parities. Dashed
(solid) line is for $\protect\sqrt{s}=0.5(1)$ TeV energy, (1),(2),(3) stand for
different $\omega^2$ values: $\omega^2=6 \cdot 10^{-3}\;TeV^{-1}$ (1);
$\omega^2=5 \cdot 10^{-4}\;TeV^{-1}$ (2); $\omega^2=5 \cdot 10^{-5}\;TeV^{-1}$ 
(3). Doubly solid line in this and next figures denotes a background 
level of this process.}
\label{fig1}
\end{figure}
\begin{figure}[h]
\epsfig{file=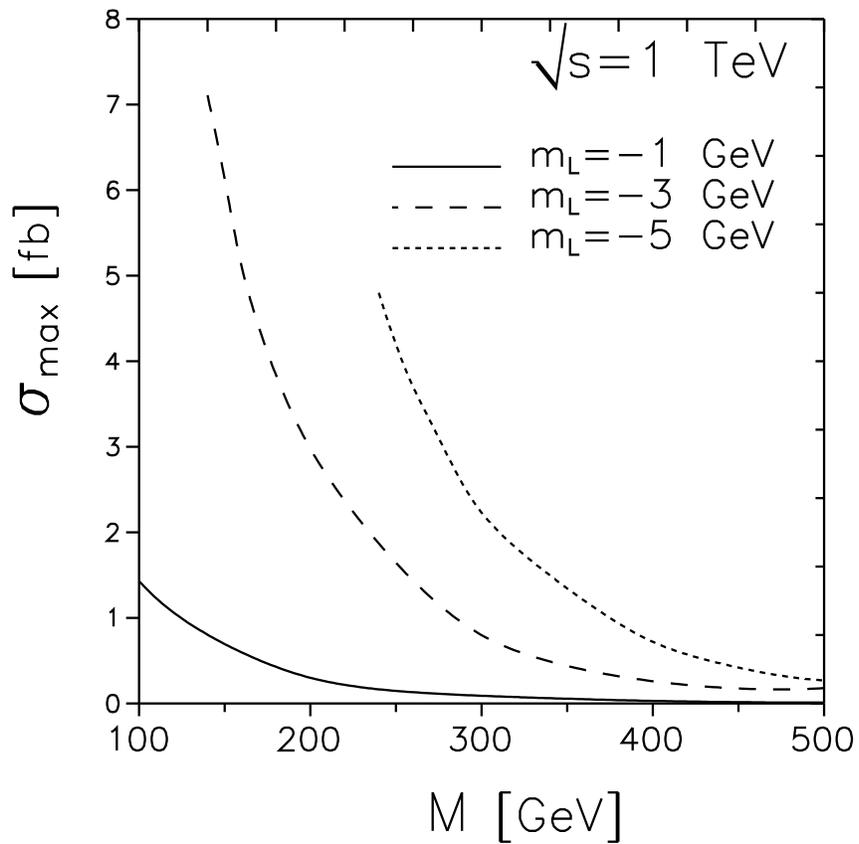,width=13cm,angle=90}
\caption{ Influence of the $m_L$ on $\sigma_{max}$ for two heavy neutrinos with
opposite CP parities  for $\protect\sqrt{s}=1$ TeV and A=5 as function of M. Only
negative $m_L$ values give substantial results in this case.}
\label{fig2}
\end{figure}
\begin{figure}[h]
\epsfig{file=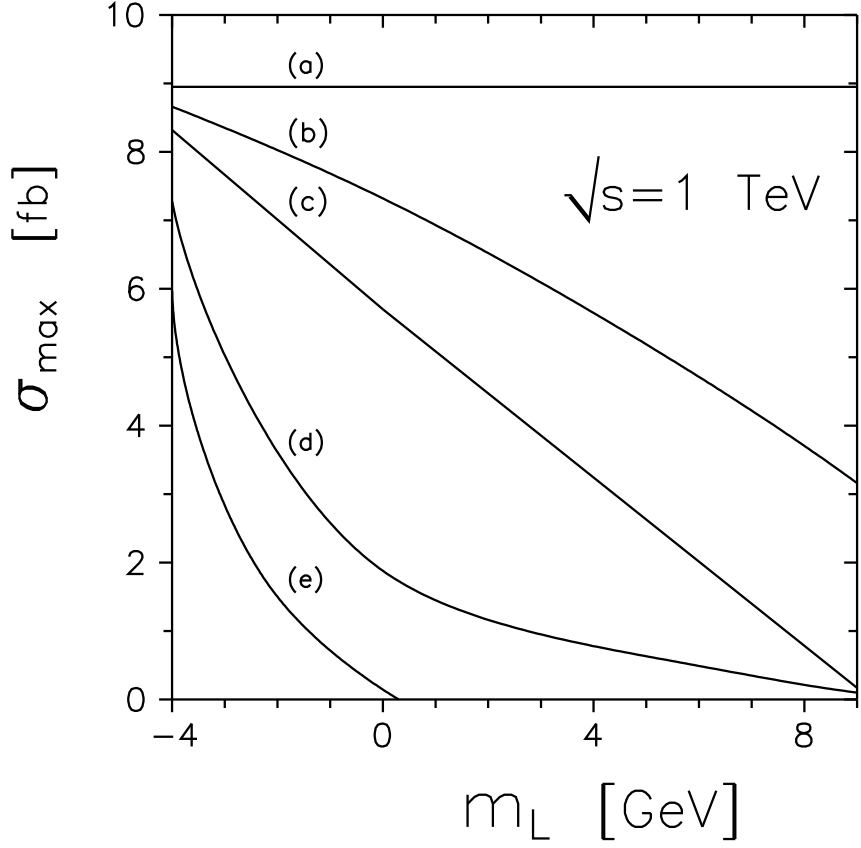,width=13cm,angle=90}
\caption{ The case of three heavy neutrinos with
$\eta_{CP}(N_1)=\eta_{CP}(N_2)=-\eta_{CP}(N_3)$. The cross section as 
function of $m_L$ for different A, B=10 and M=100 GeV
is given. (a),(b),(c),(d),(e) are
for $A=10^6,100,50,20,10$, respectively.}
\label{fig3}
\end{figure}
\begin{figure}
\epsfig{file=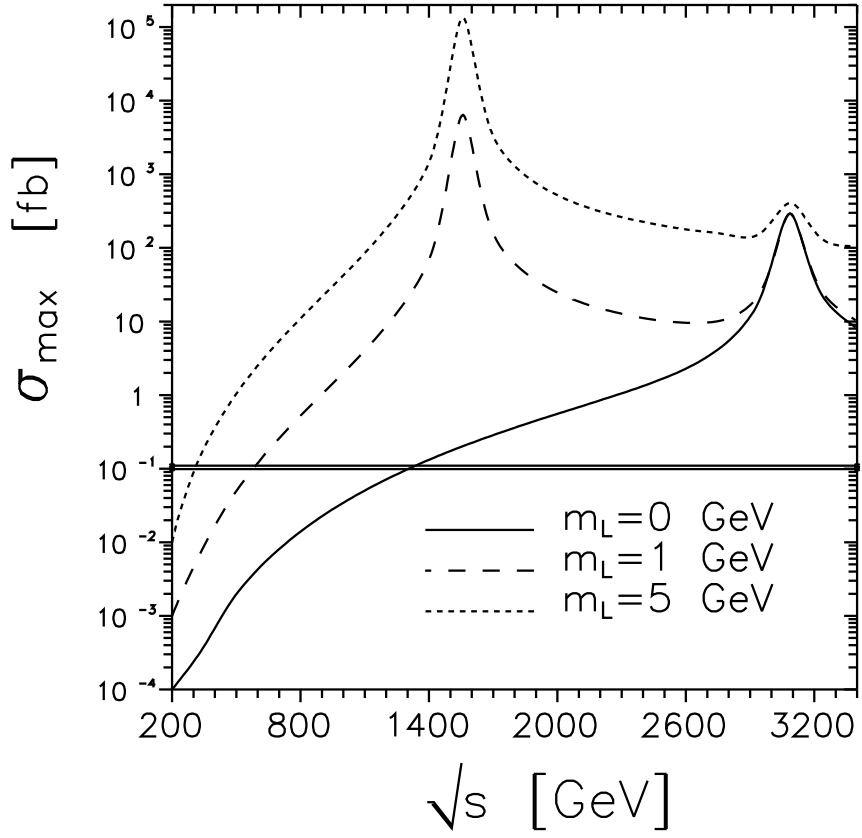,width=13cm,angle=90}
\caption{Influence of the $m_L$ parameter on the s-channel $\delta_L^{--}$ 
resonance. Solid, dashed and dotted lines are for $m_L=0,1,5$
GeV, respectively. The t and u channel contributions are calculated for
the same $\eta_{CP}$ eigenvalues of heavy neutrinos and
$\omega^2=5 \cdot 10^{-5}\;TeV^{-1}$. }
\label{fig4}
\end{figure}
\begin{figure}[h]
\epsfig{file=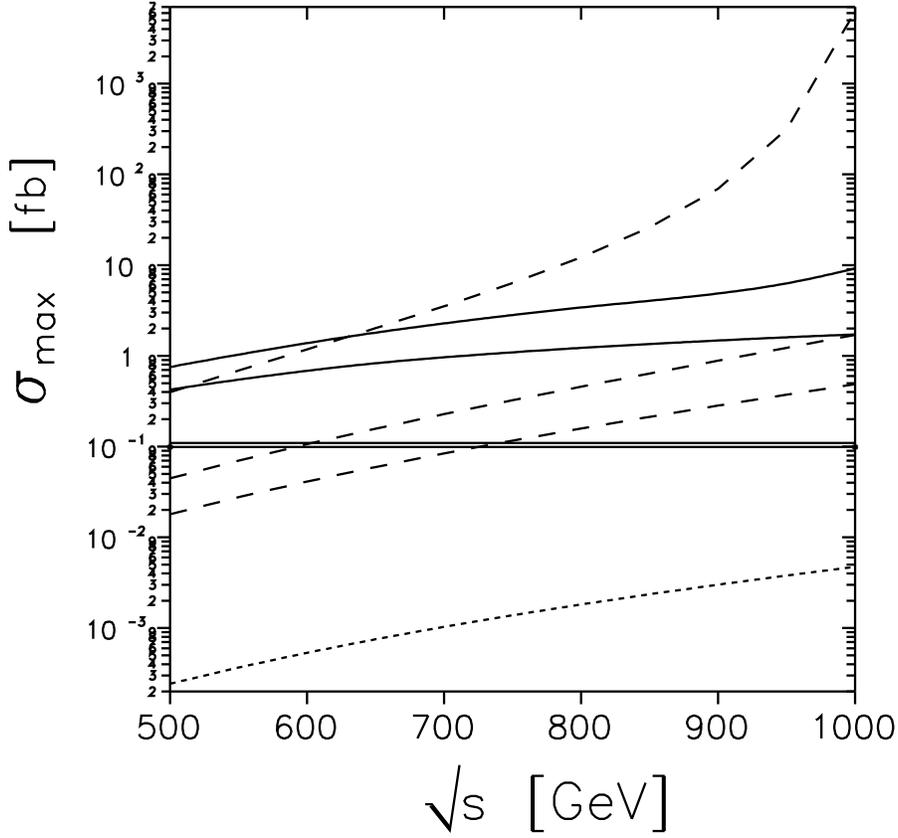,width=13cm,angle=90}
\caption{ Contribution of the $\delta_L^{--}$ resonance 
to the $e^-e^- \rightarrow W^-W^-$ process
in the range of energies of the future linear lepton
collider.
Dashed lines are for
$m_L=1$ GeV and the same CP parities of heavy neutrinos (case A), 
solid lines are for
$m_L=-1$ GeV and $\eta_{CP}(N_1)=-\eta_{CP}(N_2)$ (case B).
To show the  s channel effect we present $\sigma_{max}$ for t and u channels only
(short dashed line for the A case  and lower solid line for the B case) 
and for the full cross section with s,t and u channels 
altogether (long dashed lines for the A case  and upper solid line for the
B case). 
Long dashed lines are for $M_{\delta_L^{--}}=1000,1600,2000$
GeV, 
respectively. The upper solid line is for $M_{\delta_L^{--}}=1600$ GeV.}
\label{fig5}
\end{figure}
\end{document}